\documentclass[twocolumn,nofootinbib,prl,preprintnumbers,amsmath,amssymb]{revtex4}
\usepackage{graphicx}
\usepackage{epsfig}
\newcommand{\be}{\begin{equation}}
\newcommand{\ee}{\end{equation}}
\newcommand{\ba}{\begin{eqnarray}}
\newcommand{\ea}{\end{eqnarray}}

\newcommand{\MET}{\displaystyle{\not} E_T}
\newcommand{\mtest}{m_{\chi_1}^\text{test}}
\newcommand{\DelMET}{\Delta \displaystyle{\not} E_T}
\begin{document}
\preprint{} 
\title{Mass Measurement in Boosted Decay Chains with Missing Energy}
\author{Jay Hubisz and Jing Shao}
\affiliation{201 Physics Building, Syracuse University, Syracuse, NY 13244}
\date{\today}

\begin{abstract}

We explore a novel method of mass reconstruction in events with missing transverse momentum at hadron colliders.  In events with sizeable boost factors in the final steps of dual multi-stage decay chains, the missing energy particles may each be approximately collinear with a visible standard model particle, spanning a narrow ``MET-cone." We exploit this collinear approximation, when applicable, to reconstruct the masses of exotica.
\end{abstract}
\maketitle

{\bf {\em Introduction ---}}
The start of the Large Hadron Collider (LHC) at CERN gives hope to the discovery of long-anticipated TeV scale new physics. In many of these scenarios, discrete symmetries are motivated by the requirement that higher dimensional operators which contribute to processes such as weak gauge boson form factors and baryon number violation must be highly suppressed. Such discrete symmetries may simultaneously be responsible for stabilizing the dark matter component of our universe. The typical collider signatures of such discrete symmetries are characterized by events with large amounts of missing transverse momentum. The undetected particles in such events, the lightest particle charged under the discrete symmetry, complicate the reconstruction of the masses of new exotica. This is particularly the case at hadron colliders, where the initial parton momenta are unknown. There has been recent substantial progress in mass measurement in such scenarios over the past few years (see \cite{Barr:2010zj} for a recent review, and complete citation list). Most of these methods relies on the kinematics of the events and fall into several broad categories: invariant mass endpoint methods~\cite{Hinchliffe:1996iu,Bachacou:1999zb,Allanach:2000kt}, mass relation/polynomial methods~\cite{Nojiri:2003tu,Kawagoe:2004rz,Cheng:2007xv,Cheng:2008mg}, $M_{T2}$-like methods~\cite{Lester:1999tx,Barr:2003rg,Cho:2007qv}, and various combinations of them. The existence of multiple methods is crucial, providing complementary techniques for extracting information about the underlying physics model.

In this letter, we explore a conceptually new method of mass determination that is particularly useful when a decay chain terminates with the disintegration of a relatively boosted exotic particle to the lightest exotic plus a visible standard model (SM) particle.~\footnote{From here on, we use the semantics of supersymmetry, and refer to the lightest particle charged under the discrete symmetry as the LSP, and the next lightest as the NLSP. However, our method is generic, and applies to all TeV scale physics in which the exotica carry a conserved $Z_2$ charge under which the SM fields are neutral.}  Such events are characteristic of models in which pair produced color charged exotica are quite heavy, and in which there is simultaneously a small amount of phase space available for the NLSP decay. As a motivator for such scenarios, in the MSSM, the LEP II bound on the higgs mass prefers a large top squark mass, or in general TeV scale squark masses in models with minimal flavor violation. On the other hand, the neutralinos and charginos can naturally be light, near the weak-scale and with relatively small mass splitting.

\begin{figure}[htbp]
\begin{center}
   \includegraphics[width=2.in]{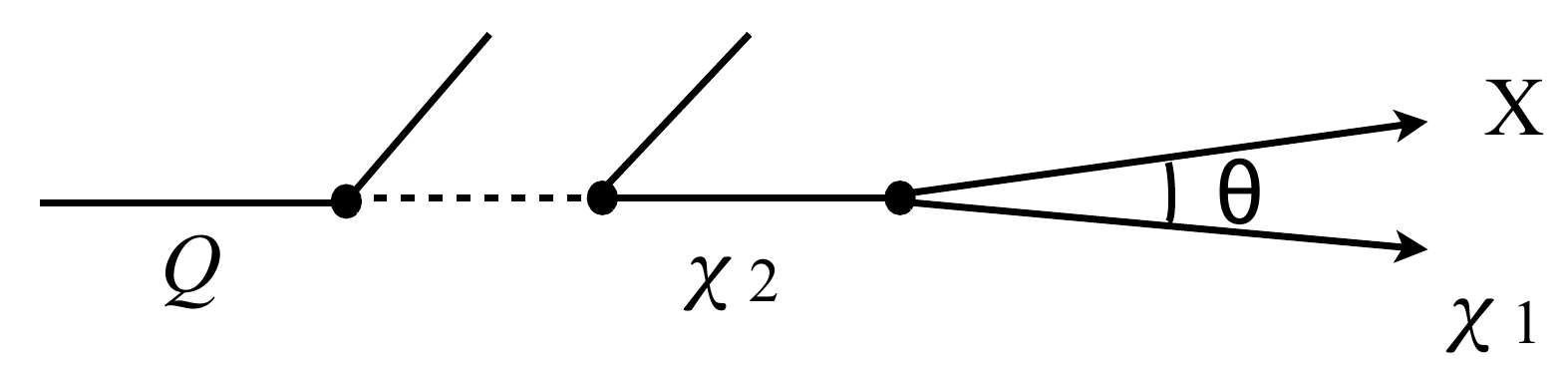}
\caption{A schematic picture of a boosted decay chain.}
\label{decaychain}
\end{center}
\vspace{-0.6cm}
\end{figure}

We consider production of a generic new heavy particle $Q$ in a collider, which decays in the following way: $Q \rightarrow  \cdots \chi_2 \rightarrow   \cdots  \chi_1 X $.  Here $\chi_{1,2}$ are the LSP and NLSP respectively, while $X$ is a SM particle. The set of dots represents a multiplicity of SM states arising from the intermediate stages of the decay chain. This rather general decay topology is shown in Fig.~\ref{decaychain}. If the mass difference between $Q$ and $\chi_2$ is large, $m_Q -m_{\chi_2}\gg m_{\chi_2}$, the daughter particle $\chi_2$ will often be highly boosted. 

We point out that for a given $X$ momentum configuration, there is a kinematic boundary for the missing momentum.  In the case of a boosted decay chain, the total missing momentum is constrained to lie in a narrow cone around the total $X$ momentum, which we call the MET-cone. This observation is the central focus of this letter, and motivates the construction of a simple variable which contains information about the mass spectrum. We begin with a brief discussion of collinearity and its dependence on the mass parameters. 
~\footnote{There is a recent study \cite{Cho:2010vz} which also uses boosted decays for mass measurement, but in a quite different way.}

{\bf {\em Collinearity and the MET-Cone ---}}
The kinematics of two-body $\chi_2$ decays are straightforward. We take the $\chi_2$ particle to have relativistic boost factor $\gamma$ and velocity $\beta$ in the lab frame. The angle $\theta_{\chi_2 X}$ between the visible particle $X$ and the direction of motion of the parent $\chi_2$ is then given by 
\begin{equation}
\tan\theta_{\chi_2 X} = \frac{ \beta^X_0}{\gamma} \left( \frac{\sin \theta_0}{ \beta^X_0 \cos \theta_0 + \beta } \right).
\end{equation}
where $\theta_0$ is angle between $X$ and $\chi_2$ in the rest frame of $\chi_2$, and $\beta_0^X$ is the velocity of $X$ in the rest frame of $\chi_2$.  The angle takes on values in the range (for $\beta_0^X <\beta$):
\begin{equation}
0 \le \tan\theta_{\chi_2 X} \le \frac{\beta_0^X}{\gamma \beta}\frac{1}{\sqrt{1-(\beta_0^X/\beta)^2}} \xrightarrow{ \gamma \gg 1 } \frac{\beta_0^X \gamma_0^X}{\gamma}.
\label{eq:maxangle}
\end{equation}
The velocity $\beta_0^X$ is a function of the masses of the three particles involved, and it characterizes the allowed phase space of the $\chi_2$ decay.
The angle $\theta_{\chi_2 \chi_1}$ can be obtained by exchanging $m_X$ with $m_{\chi_1}$ in the above equations. A collinear configuration is achieved with a large $\gamma$, and with narrow phase space for the $\chi_2$ decay. 

The $\chi_2$ boost factor is determined by several variables. As a simple example, we consider a heavy exotic, with mass $m_Q$ which decays to a massless SM particle (e.g. a jet) and the NLSP, with mass $m_{\chi_2}$. 
For $m_Q\sim 2$~TeV and $m_{\chi_2}\sim 200$~GeV, a boost factor of $\gamma = 5$ is achieved in the rest frame of $Q$. However, at a hadron collider the $Q$ particle will be produced with some transverse as well as longitudinal momentum, providing a distribution of boost factors. In addition, in multistage decay chains, the typical boost factors will depend on the mass spectrum of particles participating in the cascade. 

The boost factors of $X$ and $\chi_{1}$ in the lab frame are given by 
\begin{eqnarray}
\gamma_{\chi_1,\, X} = \gamma \, \gamma^{\chi_1, \, X}_{0} (1 \pm \beta\, \beta^{\chi_1,\, X}_{0} \cos\theta_{0}) \label{boost_chiX}
\end{eqnarray}
The magnitudes of the 3-momenta of $X$ and $\chi_{1}$ in the lab frame can be written as $p_{\chi_1}= \gamma_{\chi_1} \beta_{\chi_1} m_{\chi_1}$ and $p_{X}= \gamma_{X} \beta_{X} m_{X}$.

\begin{figure}[htbp]
\begin{center}
   \includegraphics[width=3.in]{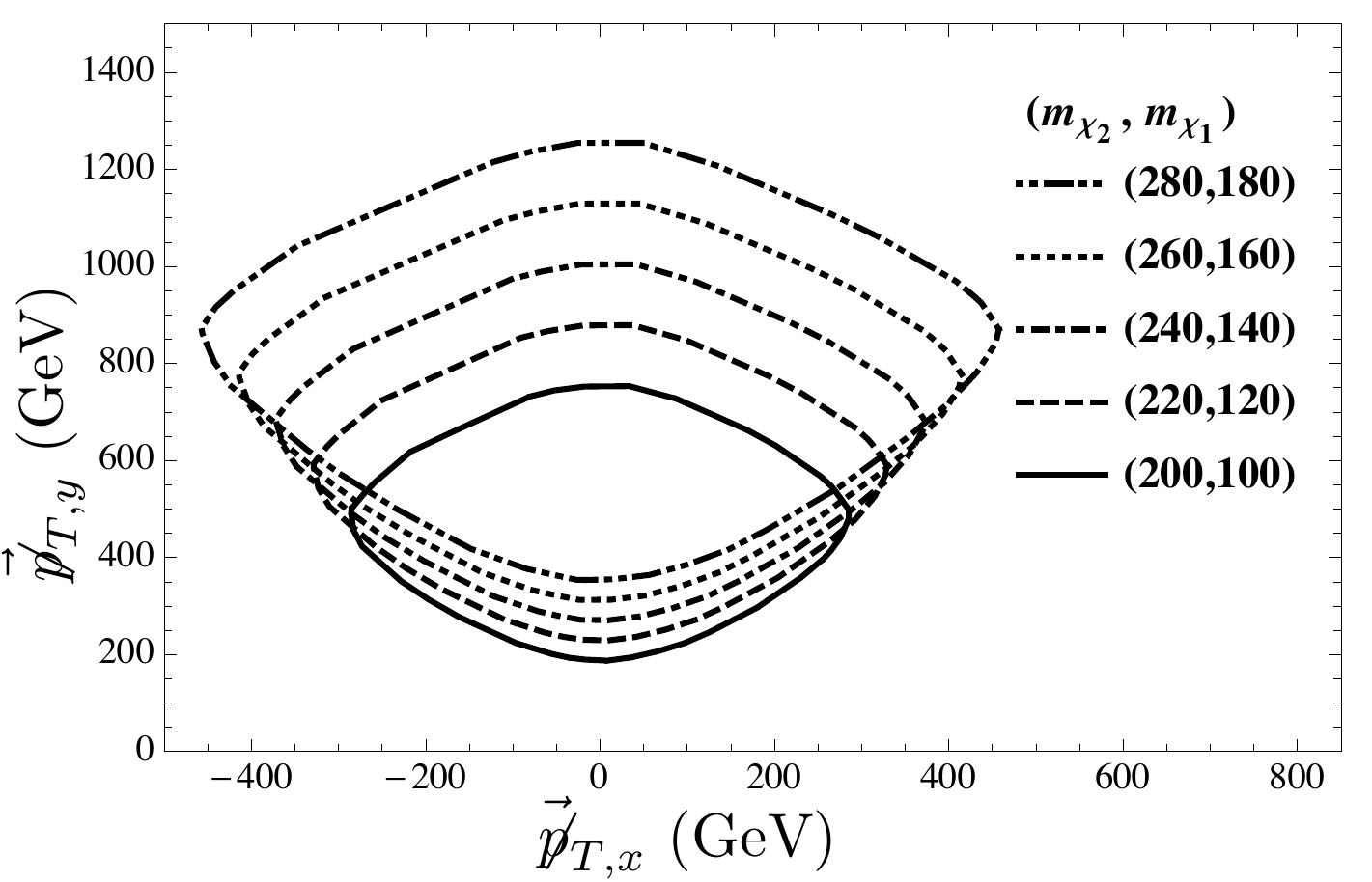}
\caption{In this figure we display the MET-cone boundary for different NLSP and LSP masses (keeping the mass splitting constant).  The total $\MET$ vector must lie within the boundary for particular choices of exotica masses.  The $Z$ bosons are in a configuration where each has boost factor $5$, both lie in the transverse plane, and are separated by a $90$ degree angle.}
\label{fig:collinearscatter}
\end{center}
\vspace{-0.8cm}
\end{figure}

Eq.'s~(\ref{eq:maxangle},\ref{boost_chiX}), define a kinematic boundary on the contribution of one $\chi_1$ to the total $\displaystyle{\not} E_T$.
These kinematic endpoints persist when there are two $\chi_1$ particles in a single event. We illustrate this in Figure~\ref{fig:collinearscatter}, where we display the region allowed for the total $\MET$ vector in each event for given NLSP and LSP masses.  We assume an event topology where all $\displaystyle{\not} E_T$ arises from two $\chi_1$ particles (i.e. there are no neutrinos in the event), and that each of the two decay chains terminates with the NLSP decaying to a $Z$-boson plus the LSP. We restrict to a particular configuration of $Z$ momenta.  The $y$-axis reflects the component of the $\displaystyle{\not} E_T$ vector parallel to the total transverse $Z$-momentum while the $x$-axis displays the remaining $\displaystyle{\not} E_T$ vector component. As expected, the total $\displaystyle{\not}E_T$ vector is correlated with the $Z$-momenta, with kinematic boundaries determined by the mass spectrum of the underlying theory.

{\bf {\em The $m^\text{test}_{\chi_1}$ variable ---}}
We utilize the collinear limit to inspire a test variable whose distribution yields the masses of the exotica. In the case of small mass splitting, $\Delta m = m_{\chi_2} - m_{\chi_1} - m_X$, the decay products are not significantly relativistic in the rest-frame of the parent, $\chi_2$. Thus in the lab frame, where the $\chi_2$ has relativistic velocity, the boost factors of all three particles are nearly the same, and the $\chi_1$ particles are closely aligned with the $X$-momenta. With this scenario as our motivation, we define ``test" missing 3-momenta as 
\begin{equation}
\vec{\displaystyle{\not}p }_{\text{test}}^{~a,b} \equiv \vec{p}_{X}^{~a,b} \frac{m^{\text{test}}_{\chi_1}}{m_X},
\end{equation}
with $m^{\text{test}}_{\chi_1}$ defined for each event by minimization of the following quantity:
\begin{equation}
\Delta {\displaystyle{\not} E}_T^2 (m^{\text{test}}_{\chi_1}) =  \left| \vec {\displaystyle{\not} p}_{\text{test}}^{~T, \text{total}}- \vec {\displaystyle{\not} p}_{\text{exp}}^{~T} \right|^2.
\end{equation}
This is an analytic procedure, as this formula is quadratic in $m^\text{test}_{\chi_1}$. The minimization results are given by $m^{\text{test}}_{\chi_1}= m_X \vec{\displaystyle{\not}p}_{T,y}/p_{X,T}^{\text{tot}}$ and $\DelMET^{\text{min}}/\MET= \left | \vec{\displaystyle{\not}p}_{T,x}/\displaystyle{\not}E_T\right|$.  The variables $\mtest$ and $\DelMET^{\text{min}}/\MET$ rescale respectively the $y$ and $x$ components of the $\MET$ vector event-by-event.

\begin{figure}[htbp]
\begin{center}
   \includegraphics[width=2.5in]{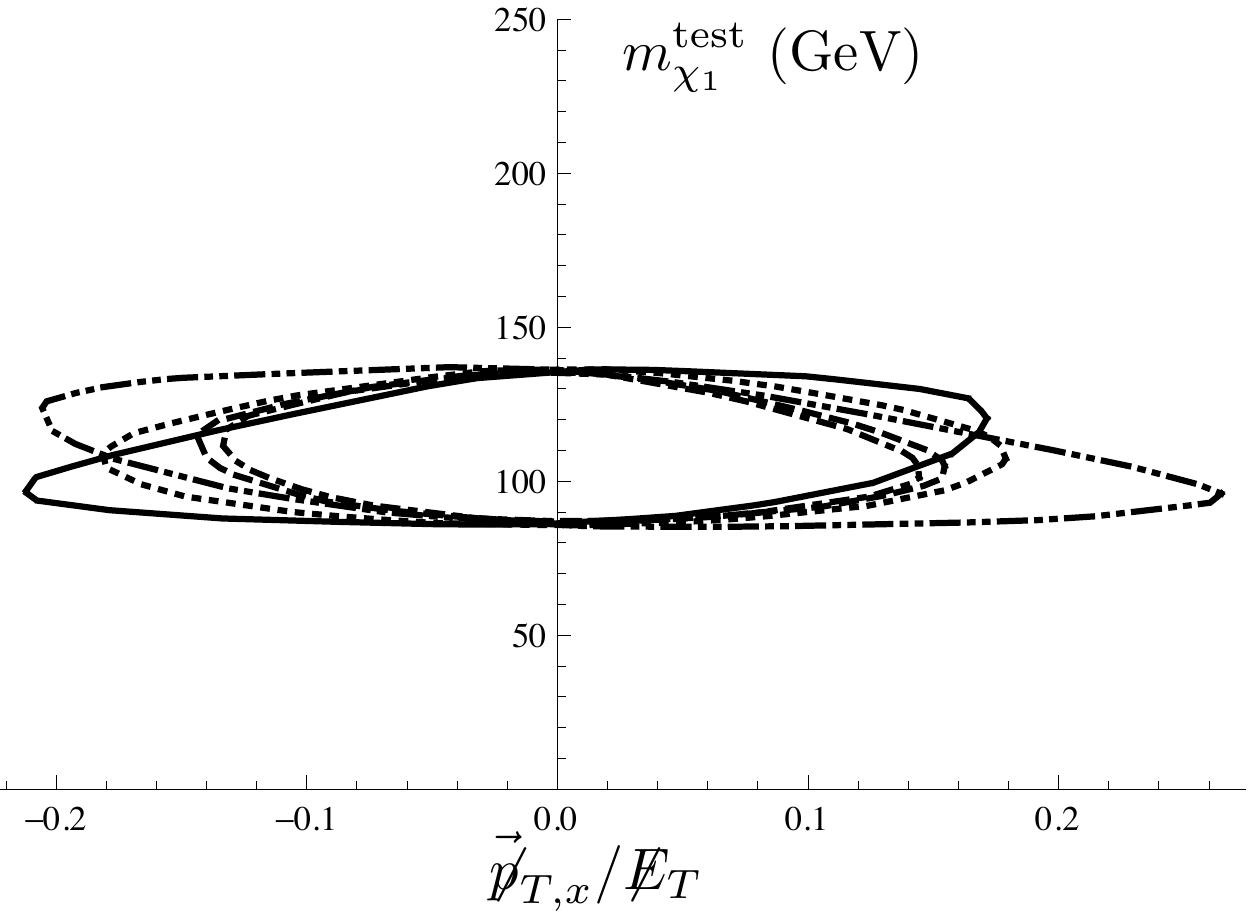}\vspace{0.1cm}
   \includegraphics[width=2.5in]{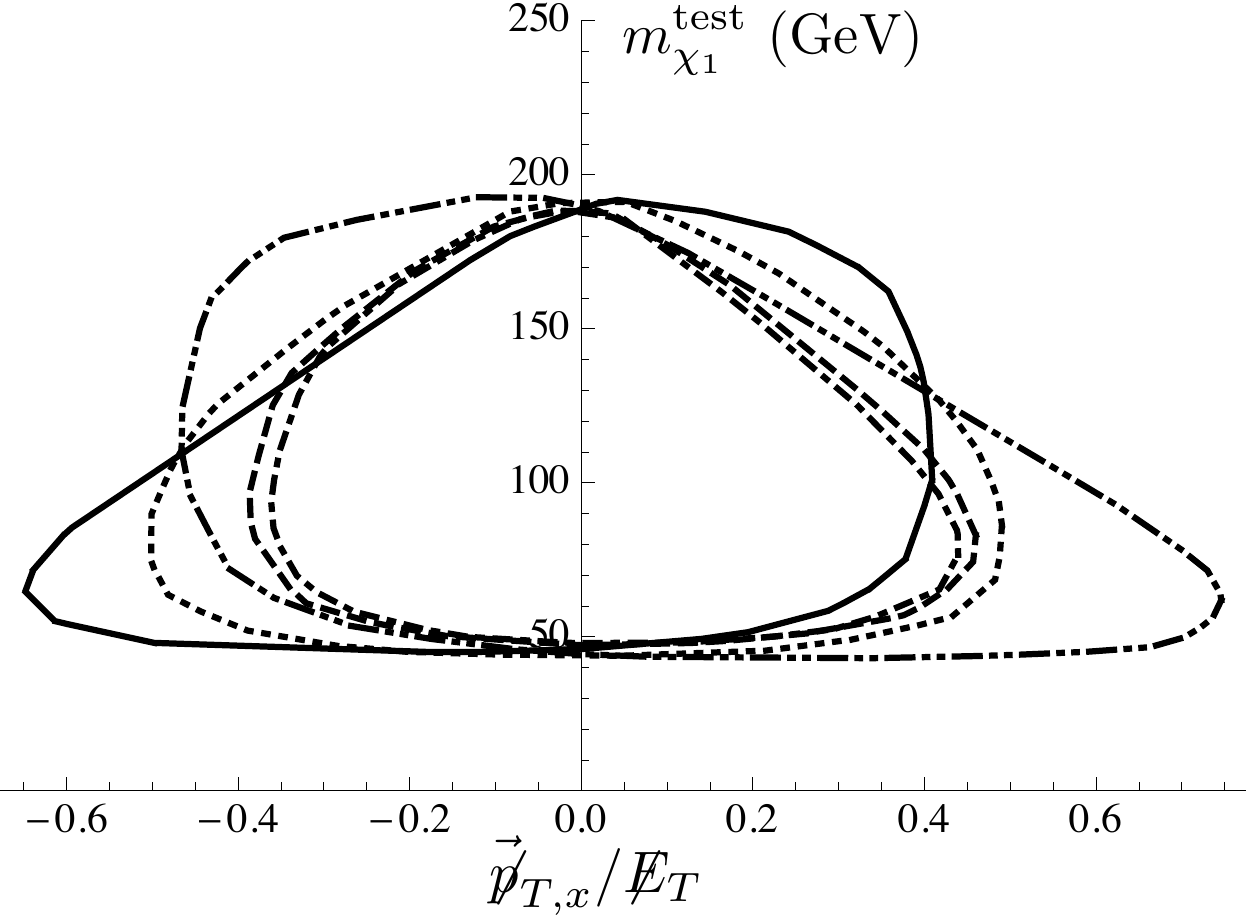} 
\caption{In these two plots, we compare the rescaled MET-cones of two scenarios with decay $\chi_2 \rightarrow \chi_1 Z$, with identical $\chi_2$ mass ($200$~GeV). In the top (bottom) plot, we take $\Delta m = m_{\chi_2} - (m_{\chi_1} + m_Z )= 1~\text{GeV} (\text{10}~\text{GeV})$. Different contours correspond to randomly chosen Z-momentum configurations.  Note that $m^{\text{test}}_{\chi_1}$ has two endpoints in the small $\DelMET^{\text{min}}/\MET$ region. In addition, the very small mass splitting scenario (top) predominantly occupies a much narrower range of $\DelMET^{\text{min}}/\MET$. }
\label{fig:mtestscatter}
\end{center}
\vspace{-0.5cm}
\end{figure}

\begin{figure}[htbp]
\begin{center}
   \includegraphics[width=2.5in]{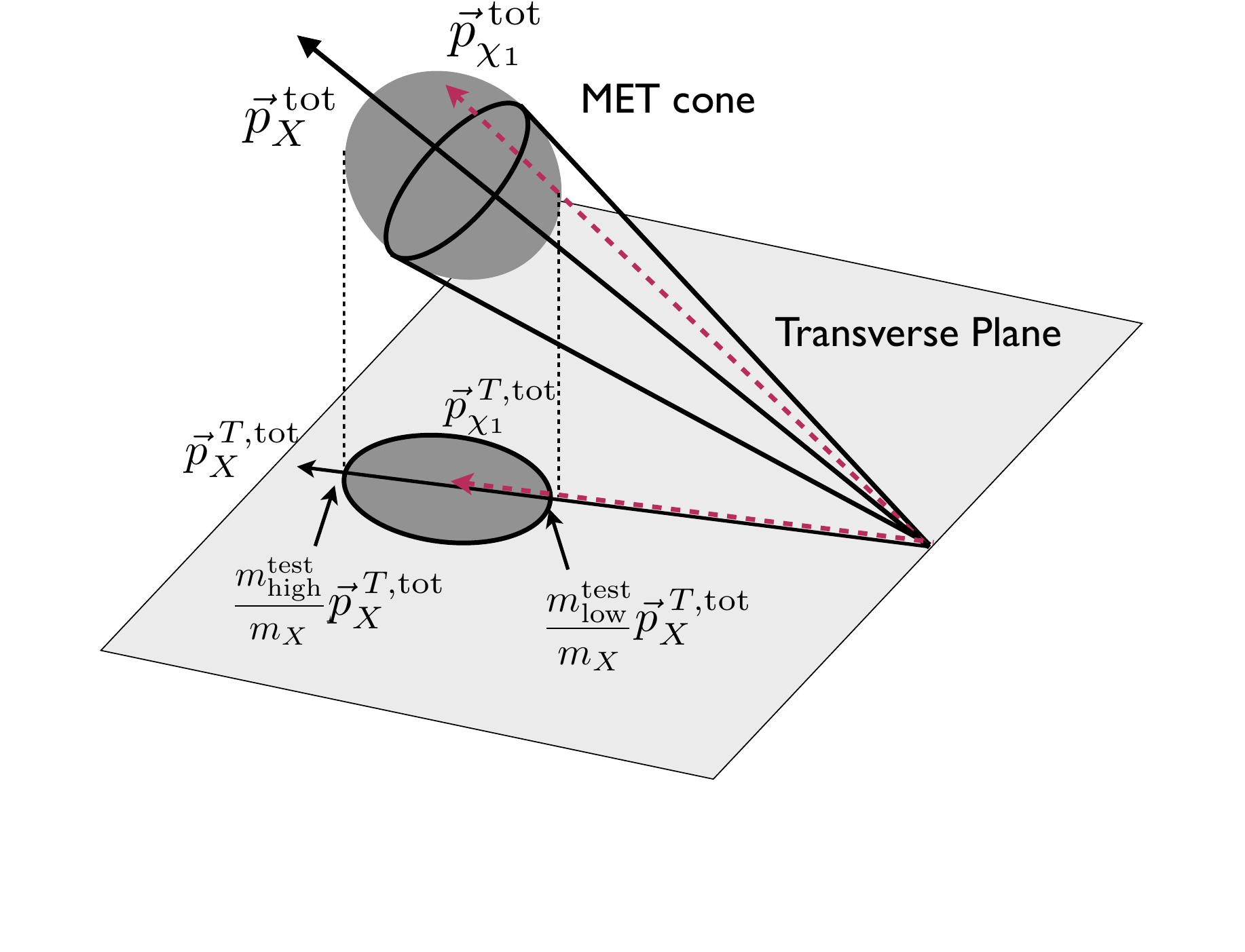}
\caption{Illustration of the MET cone and its relation to the $m_{\chi_1}^{\text{test}}$ variable. }
\label{fig:metcone}
\end{center}
\vspace{-0.5cm}
\end{figure}

As a means of quality control, we veto signal events in which  $\Delta {\displaystyle{\not} E}_T^\text{min}/ {\displaystyle{\not} E}_{T} > \epsilon $, for some sufficiently small $\epsilon$. The efficiency of such a cut is itself a rough measure of the mass splitting. In Figure~\ref{fig:mtestscatter}, we show rescaled MET-cones in the $m^{\text{test}}_{\chi_1}$ vs $\Delta {\displaystyle{\not} E}_T^\text{min}/ {\displaystyle{\not} E}_{T}$ plane for both a scenario where the mass splitting is very small, and another where an $\mathcal{O} (10\%)$ splitting is assumed. From these figures, one observes that $m^{\text{test}}_{\chi_1}$ has two approximate endpoints in the small $\Delta {\displaystyle{\not} E}_T^\text{min}/ {\displaystyle{\not} E}_{T}$ region, where the cones all intersect the $\mtest$ axis. These endpoints are a manifestation of the MET-cone boundaries when projected onto the transverse plane, as illustrated in Figure~\ref{fig:metcone}.

The limit $\epsilon \rightarrow 0$ is equivalent to an alignment condition on the momenta:
$\vec p^{\; T, \text{total}}_{\chi_1} \rightarrow \vec p^{\; T, \text{total}}_X \, m^{\text{test}}_{\chi_1}/m_{X}$. In this limit, one can express $m^{\text{test}}_{\chi_1} $ in terms of the measurable parameters of the event. The result can be written as an expansion in the angular separation between $X$ and $\chi_1$ for both sides of the decay chain, $\theta^{a,b}$, in the near-collinear case. In a configuration where both $X$'s are in the transverse plane, we obtain a relatively clean result:
\begin{eqnarray}
m^{\text{test}}_{\chi_1} &\approx &m_{\chi_1}\frac{\gamma^{\chi_1}_{0}}{\gamma^{X}_{0}} \frac{1 + \beta\, \beta^{\chi_1}_{0}\cos\theta^a_{0}}{1 - \beta\, \beta^{X}_{0} \cos\theta^a_{0}} \nonumber\\ 
&\times & \left(1 - \cot\theta^X_{ab} \cos\phi^a \theta^a + \csc\theta^X_{ab} \cos\phi^b \theta^b \right),\label{mtest-approx}
\end{eqnarray}
where $\beta$ and $\gamma$ refer to the NLSP in the a-chain, and $\beta_0^X \approx \beta_0^{\chi_1}$ is used. Here ($\theta^{a}$, $\phi^{a}$) are the spherical coordinates of $\vec p_{\chi_1}^{\; a}$ in the lab frame where the $z$-axis is along the $\vec p_{X}^{\;a}$; $\theta^X_{ab}$ is the angle between $\vec p_{X}^{\; a}$ and $\vec p_{X}^{\; b}$. At zeroth order in the $\theta^{a,b}$ expansion, the endpoints of $m^{\text{test}}_{\chi_1}$ are given by: 
\begin{eqnarray}
m^{\text{test}}_{\pm } \approx m_{\chi_1}\frac{\gamma^{\chi_1}_{0}}{\gamma^{X}_{0}} \frac{ 1 \pm \beta^{\chi_1}_{0}}{1 \mp \beta^{X}_{0}}\label{endpoint-mtest}, 
\end{eqnarray}
which are achieved when $\theta^a_{0}= 0$ and $\pi$ respectively, i.e. $\chi_1$ moving forward or backward along the $\chi_2$ boost direction in its rest frame.  The range of $m^{\text{test}}_{\chi_1}$ becomes smaller as the phase space for the $\chi_2$ decay shrinks. In the relativistic limit, $\beta \sim 1$, the endpoint positions $m^{\text{test}}_{\pm}$ approximately determine the unknown masses $m_{\chi_1}$ and $m_{\chi_2}$.

Now let us discuss the non-collinear corrections to the endpoint positions from Eq.~(\ref{mtest-approx}). First we note that $\theta^a$ depends on $\theta^a_{0}$ and in the near-collinear case it does not shift the endpoint positions. The dominant contribution comes from $\theta^{b}$ which is independent of $\theta^a_{0}$ and gives rise to a shift of the endpoint position $\Delta m^{\text{test}}_{\pm} \approx \pm m^{\text{test}}_{\pm} \csc \theta^X_{ab}\, \theta^{b} $. Since $\theta^b$ follows a distribution determined by the kinematics of the decay, it leads to a smearing of the distribution near these endpoints. In the near-collinear case, the variation of $\theta^b$ around the central value is small, and smearing is minimal.

In the general case where the $X$'s are not in the transverse plane, one needs to project the MET cone into the transverse plane and then impose the alignment condition. One can still express the result in a collinear expansion.  The zeroth-order result remains the same as that in Eq. (\ref{mtest-approx}). However, the higher order expansion coefficients are modified by trigonometric functions expected to be of order one. 

In summary, if the endpoint positions of the $m^{\text{test}}_{\chi_1}$ distribution are measured from the data, we can find solutions for the masses of the LSP and NLSP using the relation in Eq.~(\ref{endpoint-mtest}). 

{\bf {\em Numerical Results ---}}
We now explore the effectiveness of the $m^\text{test}_{\chi_1}$ method using Monte-Carlo simulation. We consider squark pair production $\tilde q_L \tilde q_L$ followed by the decays $\tilde q_L \rightarrow q \tilde \chi_2 \rightarrow q \tilde \chi_1 Z$ in SUSY models. We consider the four spectra shown in Table~\ref{table:mass}.
\begin{table}[htbp]
\begin{centering}
   \begin{tabular}{c|ccccc}
   \hline
     Model & $m_{\chi_1}$ & $m_{\chi_2}$ & $m_{\tilde q_L}$ & $(m_{-}^{\text{test}})^{\text{theo}}$ & $(m_{+}^{\text{test}})^{\text{theo}}$\\ \hline
    $1$ &  $100$ & $200$ & $1000$  & $54.6$   & $183.2$ \\
    $2$ &  $100$ & $250$ & $1250$  & $21.6$   & $463.0$ \\
    $3$ &  $200$ & $300$ & $1000$  & $117.9$ & $339.2$\\ 
    $4$ &  $200$ & $350$ & $1250$  & $52.6$   & $761.0$\\ 
    \hline
  \end{tabular}
  \caption{The relevant masses in four SUSY models and the expected endpoints $(m_{\pm}^{\text{test}})^{\text{theo}}$. Masses are given in GeV.}
  \label{table:mass}
\end{centering}
\vspace{-0.3cm}
\end{table}
For each of these models, we simulate $20$k events of squark pair production and decay in $pp$ collisions at $\sqrt{s}=14$~TeV in MadGraph~\cite{Alwall:2007st} using the $2\rightarrow 6$ matrix element.  Events are selected according to the parton-level cuts in Table \ref{table:cuts}. The $\cos\theta^{ab}$ cut is to ensure the coefficients in the $\theta^{a,b}$ expansion are not too large, which would obscure the endpoints. For Models 2 and 4, we have slightly loosened the selection cuts in order to get better statistics near the tail of the distribution: {\textbf a)} $\epsilon =0.2$ for Model 2 and 4. {\textbf b)} $|\eta^{Z,\text{tot}}|<3.0$ for Model 4. 
\begin{table}[htbp]
\begin{centering}
   \begin{tabular}{cc|c|c|c|c|c}
    \hline
           &  $p_{T}^{Z}$~ &  ~$|\eta^{Z}|$~ & ~$\left|\eta^{Z,\text{tot}}\right|$~ & ~$\displaystyle{\not} E_T$~ & ~$\epsilon$~ & ~$\left|\cos\theta_{ab}^X\right|$  \\ \hline
            &  $>50$~GeV~ &  ~$<3.0$~ &  ~$< 1.0$~ & ~$>200$~GeV~  &  ~$0.15$~  &  ~$<0.5$  \\
    \hline
  \end{tabular}
  \caption{Event selection cuts for $m^{\text{test}}_{\chi_1}$. Here $\eta^{Z,\text{tot}}$ is the pseudorapidity of the total $Z$ momentum. }
  \label{table:cuts}
\end{centering}
\vspace{-0.3cm}
\end{table}
\begin{figure}[htbp]
\begin{center}
\includegraphics[width=3.in]{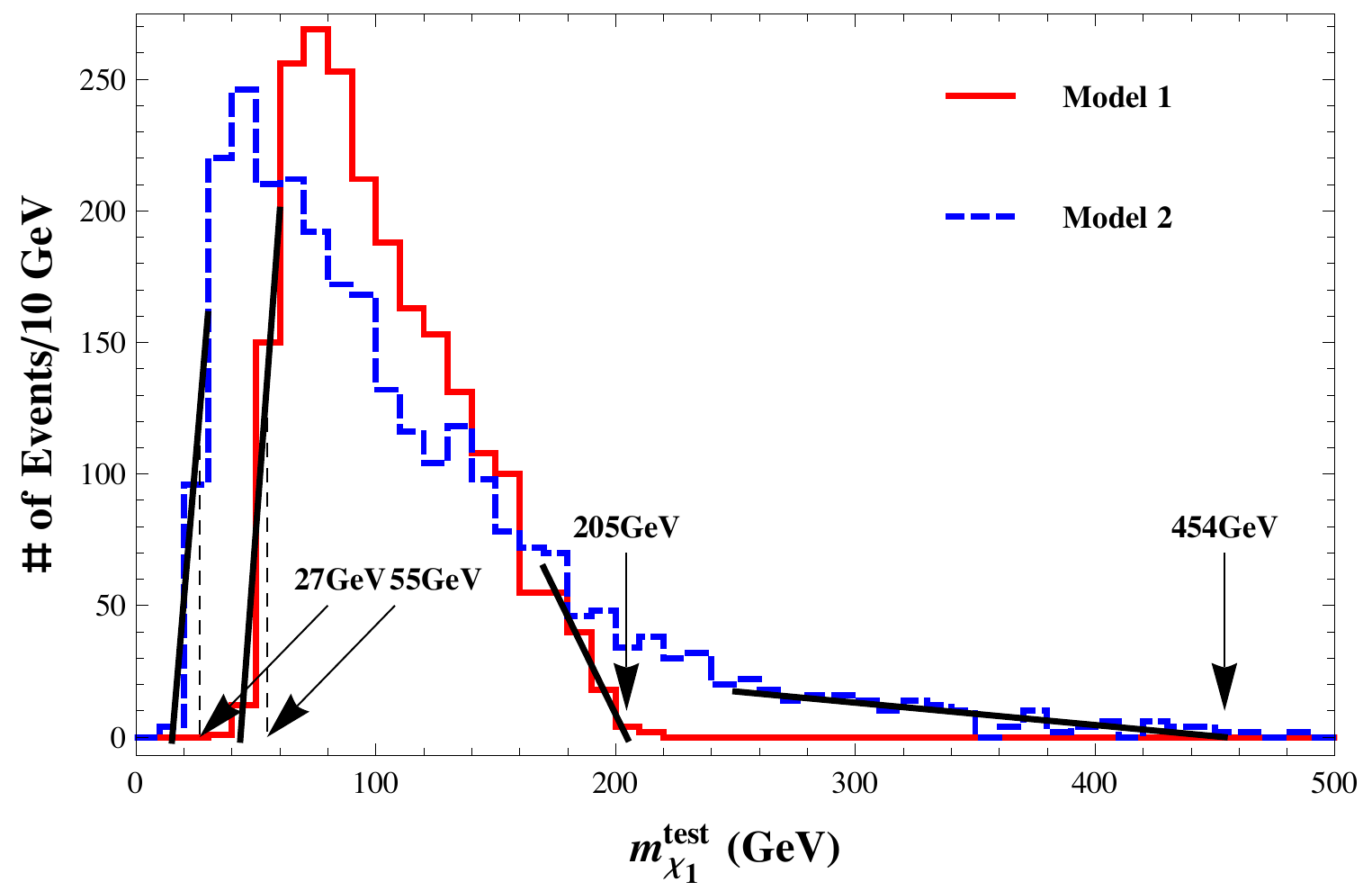}
\caption{The distributions of $m^{\text{test}}_{\chi_1}$ for Model 1 and 2. For Model 2, we have normalized the distribution by a factor of two.}
\label{mtest}
\end{center}
\vspace{-0.5cm}
\end{figure}

In Fig.~\ref{mtest}, we show the $m^{\text{test}}_{\chi_1}$ distribution for Models 1 and 2. Models 3 and 4 are very similar. 
For Model 1, we can see that the distribution is approximately a triangle with two endpoints at around $50$ and $200$~GeV. The shape of the distribution near each endpoint is slightly smeared due to deviations from collinearity.  A simple way to extract these endpoints can be achieved by a linear fit and taking the x-intercept.  This would typically under-estimate $m_{-}^{\text{test}}$ while over-estimating $m_{+}^{\text{test}}$. In our analysis, we take the position at the half maximum near the lower edge for $m_{-}^{\text{test}}$ to get a better estimation. For the upper endpoint, we use the $x$-intercept of a linear fit. More complicated fits and estimation are certainly possible. 

The results are shown in Table~\ref{table:result} together with the statistical errors. The results are consistent with the expected zeroth-order endpoints, as shown in Table \ref{table:mass}.  For Model 2, as seen in Fig.~\ref{mtest}, the upper endpoint is much less sharp than the lower one.  Estimation of its position is subject to a relatively large systematic uncertainty depending on the binning and the choice of the fitting region. Fortunately, the calculated masses are not very sensitive to the upper endpoint position. For a reasonable estimate of the $m_{+}^{\text{test}}$ in the range $400 - 500$~GeV,  the calculated masses $(m_{\chi_1}^{meas}, m_{\chi_2}^{meas})$ vary only mildly from $(103~\text{GeV}, 241~\text{GeV})$ to $(116~\text{GeV}, 264~\text{GeV})$. Therefore, even in this relatively less collinear case we still obtain a good estimate of the masses.  For all four models, the estimated endpoints and calculated masses are summarized in Table \ref{table:result}. The fitted masses are all within $\sim 10\%$ of the true masses. 
\begin{table}[t]
\begin{centering}
   \begin{tabular}{c|cccccc}
    \hline
     Model & $m_{\chi_1}$ & $m_{\chi_2}$ &  $m_{-}^{\text{test}}$ & $m_{+}^{\text{test}}$ & $m_{\chi_1}^{meas}$ & $m_{\chi_2}^{meas}$ \\ \hline
    $1$ &  $100$ & $200$  & $55\pm 2$ & $205 \pm 3 $ &  $106\pm 2$ & $208\pm 3$ \\
    $2$ &  $100$ & $250$  & $27\pm 2$ & $454\pm 20$ & $110\pm 5$ & $253\pm 5$ \\
    $3$ &  $200$ & $300$  & $112\pm 5$ & $342\pm 10$ & $195\pm 5$ & $296\pm 5$ \\ 
    $4$ &  $200$ & $350$  & $49\pm 2$ & $682\pm 16$ & $183 \pm 5$ & $329\pm 5$ \\ 
    \hline
  \end{tabular}
  \caption{Results of the measured $m_{\chi_1}^{\text{test}}$ endpoints and $m_{\chi_{1,2}}$ for four SUSY models. Masses are given in GeV. }
  \label{table:result}
\end{centering}
\vspace{-0.5cm}
\end{table}

{\bf {\em Summary and Outlook---}}
In this letter, we explored a novel method of measuring the absolute mass scale of exotic particles in events with missing energy which is inspired by boosted cascade decays.  This method uses the fact that in the boosted decay there is a limited variation in both the direction and magnitude of the total three-momentum of missing particles relative to the total three-momentum of the visible partners. The boundary of the allowed region, or the MET-cone, is determined by the mass parameters and the configuration of the visible particles. We constructed a variable $m^{\text{test}}_{\chi_1}$ which has has endpoints in the approximate collinear case. These endpoints depend on the masses involved in the final step decay, and once observed from the data, can be used to determine these masses. 

The $m^{\text{test}}_{\chi_1}$ variable works best in the collinear limit.  Given the data, the evidence of collinearity in the final step decay can be seen in various ways. First, one would see a peak at $0$ in the $\Delta {\displaystyle{\not} E}_T^\text{min}/ {\displaystyle{\not} E}_{T}$ distribution. Second, one would see well-defined endpoints in the $m^{\text{test}}_{\chi_1}$ distribution.  Once the masses of $\chi_1$ and $\chi_2$ are measured, one can find the masses of heavier exotics upstream of the NLSP decay using more standard techniques. 

We have demonstrated our method in a setup with two symmetric decay chains with a two-step cascade decay, however it applies for longer decay chains as well.  An advantage of this method is that one does not require information on all of the visible SM particles.  

The $m^\text{test}_{\chi_1}$ variable has the advantage that it is simple to calculate on an event-by-event basis.  However, it does not fully utilize the information available in the MET-cones. Developing a more effective method which takes full advantage of these kinematic boundaries is currently under investigation. 

{\bf {\em Acknowledgements ---}}
We would like to thank Yuri Gershtein, Gordon Kane, Myeonghun Park, Maxim Perelstein and Felix Yu for useful discussions. JH and JS thank Cornell University for their hospitality where part of this research was conducted. JS would also like to thank the Center for High Energy Physics at Peking University and Shanghai Jiaotong University for their hospitality where part of the research was conducted. This work of JS is supported by the Syracuse University College of Arts and Sciences. JH is supported by the Syracuse University College of Arts and Sciences, and by the U.S. Department of Energy under grant DE-FG02-85ER40237.

\end{document}